\documentclass{article}

\usepackage{arxiv}

\usepackage[utf8]{inputenc} 
\usepackage[T1]{fontenc}    
\usepackage{hyperref}       
\usepackage{url}            
\usepackage{booktabs}       
\usepackage{amsfonts}       
\usepackage{nicefrac}       
\usepackage{microtype}      
\usepackage{lipsum}		
\usepackage{graphicx}
\usepackage{amsmath}
\usepackage{bm}

\title{Parameter estimation in FACS-seq enables high-throughput characterization of phenotypic heterogeneity}

\author{{\hspace{1mm}Huibao Feng}\\
	Department of Chemical Engineering \\
	Tsinghua University \\
	Beijing, China \\
	\texttt{fhb\_14@163.com} \\
	\And
	{\hspace{1mm}Chong Zhang} \\
	Department of Chemical Engineering \\
	Tsinghua University \\
	Beijing, China \\
	\texttt{chongzhang@tsinghua.edu.cn} \\
}

\date{}

\renewcommand{\headeright}{}
\renewcommand{\shorttitle}{}

\begin{document}
	\maketitle
	
	\begin{abstract}
		Phenotypic heterogeneity is a most fascinating property of a population of cells, which shows the differences among individuals even with the same genetic background and extracellular environmental conditions. However, the lack of high-throughput analysis of phenotypic diversity has limited our research progress. To deal with it, we constructed a novel parameter estimation method in FACS-seq, a commonly used experimental framework, to achieve simultaneous characterization of thousands of variants in a library. We further demonstrated the model's ability in estimating the expression properties of each variant, which we believe can help to decipher the mechanisms of phenotypic heterogeneity.
	\end{abstract}


	\section{Introduction}
	Cells are sophisticated instruments driven by the central dogma that build varieties of lives. Although traits of a cell seem simply determined by genes and environment, phenotypic diversity was found among genetically identical cells even under the same environmental condition\cite{elowitz2002stochastic}. This phenomenon, referred to as phenotypic heterogeneity, has been a major research hotspot in quantitative biology\cite{ackermann2015functional} that reveals non-deterministic\cite{elowitz2002stochastic} and non-linear\cite{ackermann2015functional}\cite{heltberg2019chaotic} properties of intracellular processes.
	
	Hitherto, there are mainly three methods to quantitatively characterize phenotypic heterogeneity: time-lapse microscopy\cite{locke2009using}, flow cytometry\cite{newman2006single}\cite{silander2012genome} and  single-cell omics\cite{macosko2015highly}\cite{heath2016single}\cite{zenobi2013single}. However, each of them can only assay one type of cell at a time, which is time-consuming and labor-intensive when testing different variants. On the other hand, since the phenotypic heterogeneity itself served as a heritable trait\cite{ansel2008cell}, obtaining the comprehensive landscape between genetic variation and corresponding expression noise can help to deepen our understanding of the causal relationship between them. Thus, there is a critical need to design a novel method that enables high-throughput characterization of phenotypic diversity. 
	
	To address this issue, we focused on a recently developed powerful experimental framework, fluorescence-activated cell sorting and high-throughput DNA sequencing\cite{kosuri2013composability}\cite{townshend2015high} (FACS-seq, Figure \ref{fig:Figure1}A). This approach enables us to associate genetic variation with the expression of the fluorescent protein, thereby allowing quantification of the expression levels of thousands of variants in parallel. However, it is generally hard to figure out the phenotypic heterogeneity of each variant. We thought this issue may root in the underutilization of data, as one conventionally less explored phenomenon is that the phenotypic heterogeneity can affect the fraction of cells across bins when sorted based on their fluorescence intensity through FACS\cite{wang2020dynamics} (Figure \ref{fig:Figure2}B). Besides, sequencing data (Figure \ref{fig:Figure1}B) and the fluorescence intensity distribution (Figure \ref{fig:Figure1}C) of the whole library were neglected in previous works, which may lead to the loss of important information. Therefore, by taking these factors into consideration, we may have chances to derive the detailed properties of gene expression.
	
	
	In this paper, based on the empirical discovery that gene expression follows a log-normal distribution, we present a novel parameter estimation method that enables quantification of phenotypic heterogeneity in a massively parallel manner using FACS-seq data. Our model contains two parts, in the first part we apply a maximum likelihood estimation (MLE) to get the parameter estimates for frequency distribution; while for the second part, a parametric generative adversarial network\cite{1406.2661} (GAN) is ultilized to fit the overall fluorescence distribution. These two parts are assembled into an artificial neural network where parameters can be updated through back-propagation. As a result, this model achieves remarkable performances in both simulation and actual FACS-seq data, which paves the way for the study on sequence-heterogeneity association.
	
	\section{Methods}

	\subsection{General framework of FACS-seq}
\begin{figure}[h]
	\centering
	\includegraphics[scale=1.3]{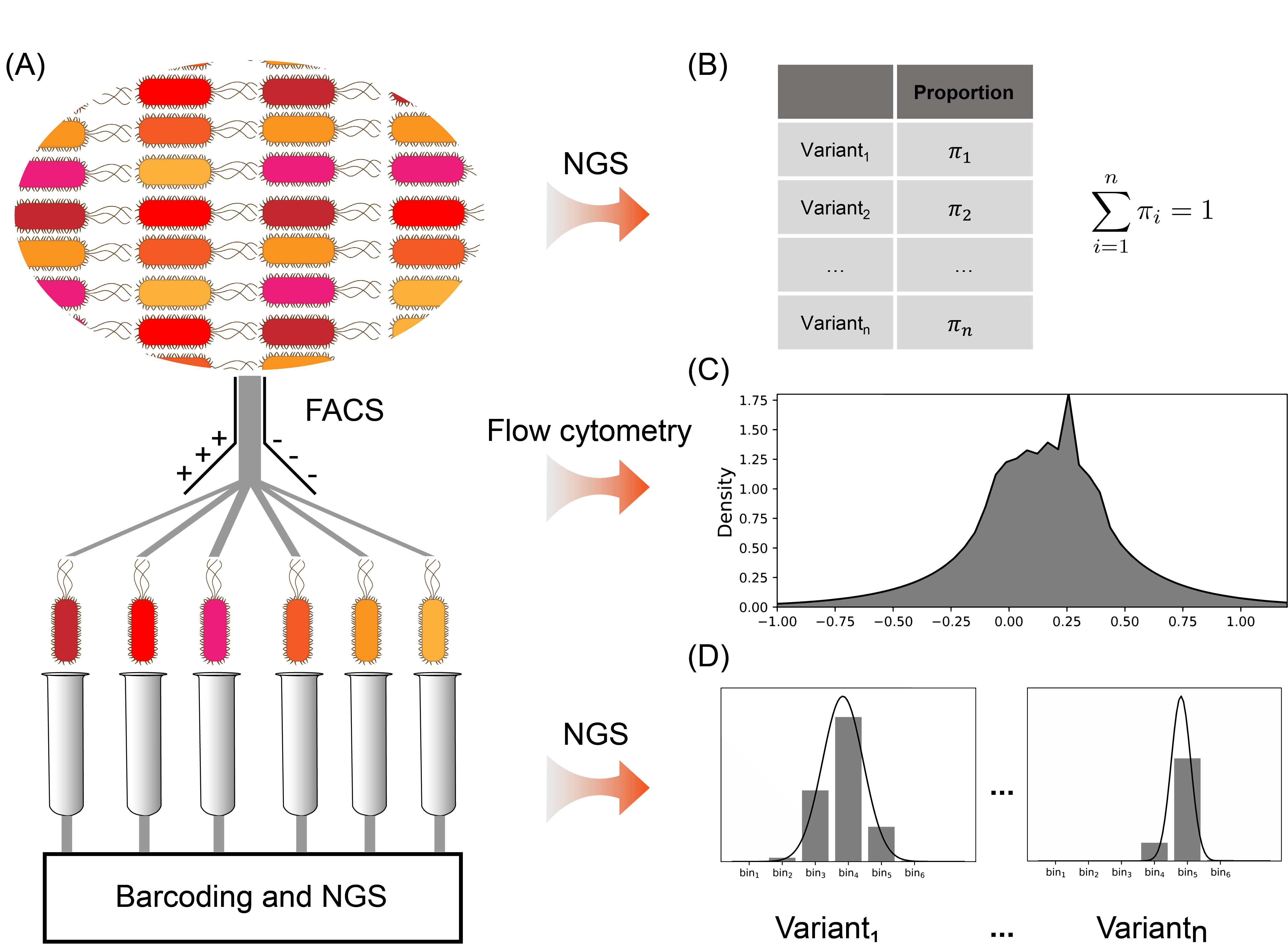}
	\caption{Experimental framework of FACS-seq. (A) Cells are encapsulated into single cell droplets and injected into the flow cytometer instrument, which are then sorted into several bins based on their fluorescence intensity. (B) Sequencing data of the whole library shows the mixing coefficient of each variant. (C) The overall fluorescence intensity distribution can be quantified by measuring tens of millions of cells. (D) The frequency distribution of each variant can be derived using the NGS data of each bin coupled with the overall distribution.}
	\label{fig:Figure1}
	\end{figure}

	FACS-seq is a method that combines two powerful high-throughput experimental frameworks together (Figure \ref{fig:Figure1}A), where the FACS part can quickly examine the fluorescence intensity ($x$, log-scaled) of millions of cells as well as separate them based on their intensity value. Hence, an overall fluorescence distribution ($p(x)$, Figure \ref{fig:Figure1}C) can be measured through it. Besides, based on the customized boundaries, cells can be sorted into several bins. Suppose in an experiment, a library with $n$ mutants are sorted into $K$ bins with boundaries $\bm{b}=\{b_0=-\infty,b_1,...,b_{K-1},b_K=+\infty\}$, where a cell with fluorescence intensity of $x$ will be screened into $bin_k$ if $b_{k-1} < x \le b_k$. The NGS part enables quantification of the abundance of each variant in a population, thus the proportion of arbitary variant $i$ ($\pi_i$, Figure \ref{fig:Figure1}B) in the library can be obtained. Moreover, based on the overall distribution and the sequencing data of each bin, we can further derive the frequency distribution of arbitary variant $i$ in bin $k$ (denoted by $f_{ik}$, see Results), which satisfies
	\begin{gather}
	\sum_{k=1}^K f_{ik}= 1, \ i=1,2,...,n.
	\end{gather}

	\subsection{Gene expression follows a log-normal distribution}
\begin{figure}[h]
		\centering
		\includegraphics[scale=1.3]{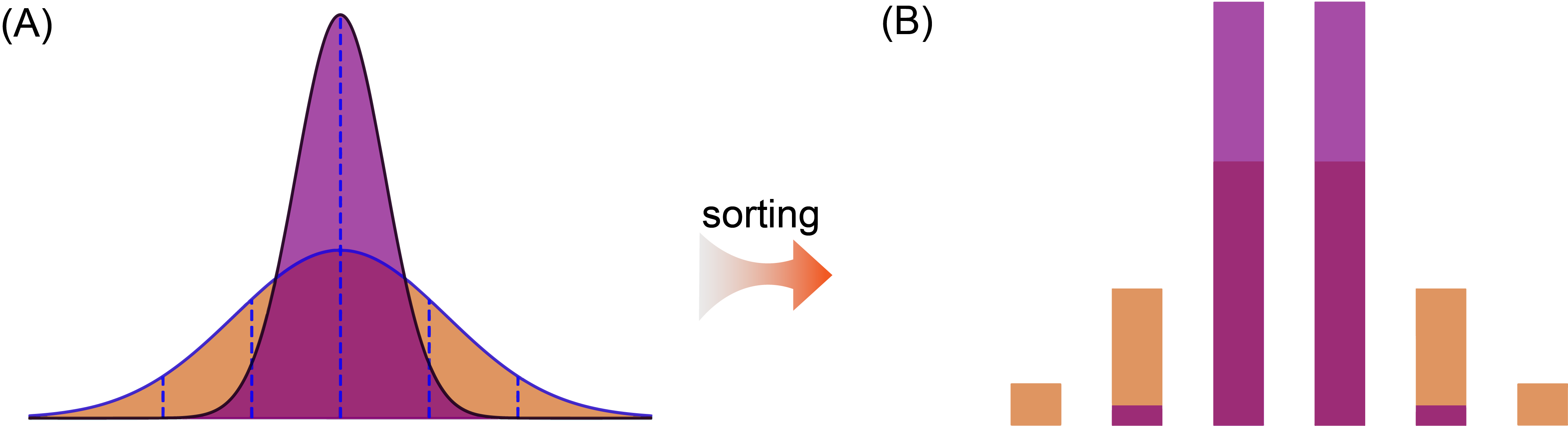}
		\caption{Different expression variations leads to varying frequency distributions. (A) Gene expression follows a log-normal distribution. (B) Given boundaries, cells are sorted into different bins. The fraction across bins is determined by the shape of gene expression distribution.}
		\label{fig:Figure2}
	\end{figure}

	The key point of describing the phenotypic heterogeneity is to find an effective statistical measurement to characterize the fluctuation in gene expression, where the coefficient of variance is commonly used in empirical studies. To derive it, the knowledge of gene expression distribution is helpful that enables precise estimation of corresponding parameters. Fortunately, plenty of works have shown that the gene expression follows a log-normal distribution\cite{beal2017biochemical} (Figure \ref{fig:Figure2}A), which can be represented using two parameters, mean and standard deviation. Suppose a library contains $n$ variants, the log-scaled expression level of variant $i$ (denoted by $x_i$) would follow a normal distribution with parameter set $(\mu_i,\sigma_i^2)$:
	
	\begin{gather}
	x_i \sim N(x_i|\mu_i, \sigma_i^2), \ i=1,2,...,n.
	\end{gather}

	\subsection{Model architecture}
\begin{figure}[h]
		\centering
		\includegraphics[scale=1.3]{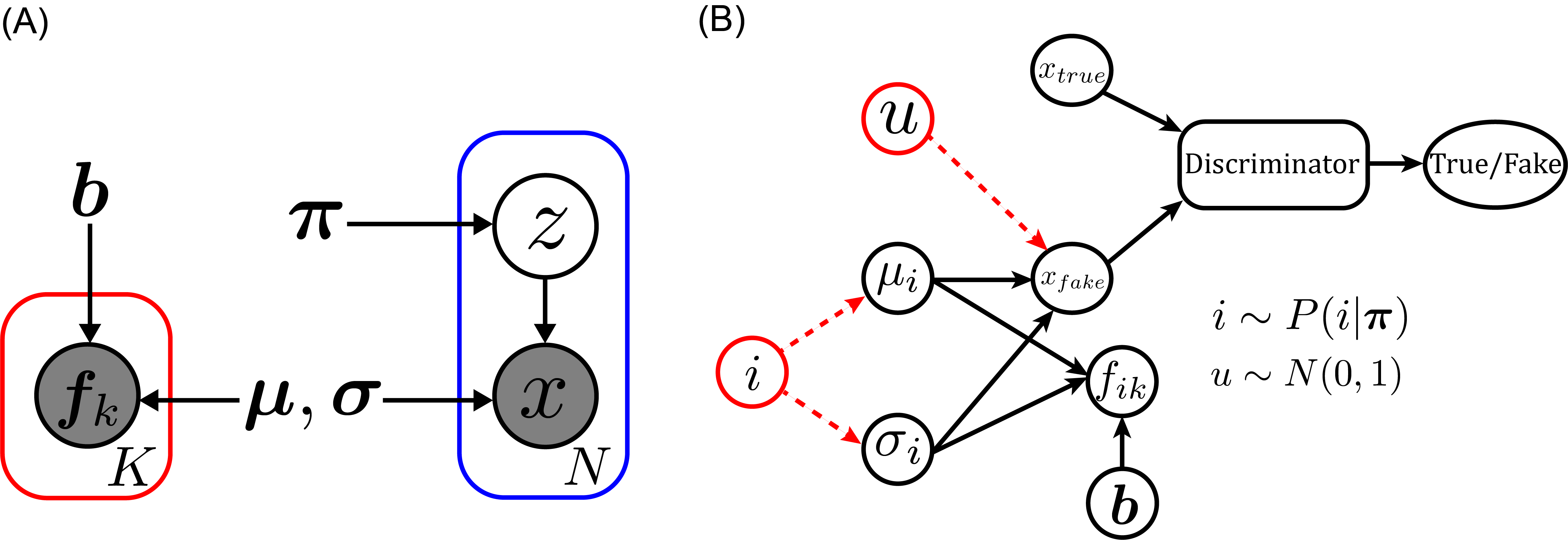}
		\caption{Architecture of the parameter estimation model. (A) Graphical representation of the model, suppose $N$ cells are measured. (B) The whole structure of the model, estimates of parameters are derived through backpropagation.}
		\label{fig:Figure3}
	\end{figure}

	According to the above descriptions, the probability of sorting variant $i$ into $bin_k$ is
	\begin{gather}
	P(x_i \in bin_k) = \int_{b_{k-1}}^{b_k} N(x_i | \mu_i, \sigma_i^2) dx_i.
	\end{gather}
	While the observed fraction of variant $i$ in $bin_k$ is
	\begin{gather}
	P(x_i \in bin_k) = f_{ik}.
	\end{gather}
	Therefore, a straightforward idea is to apply MLE to get corresponding parameter estimates, which is equivalent to minimize the following cross entropy loss function, whose derivatives with respect to parameters can be calculated using error function:
	\begin{gather}
	loss_{frac} = -\sum_{i=1}^n \sum_{k=1}^K \left( f_{ik} \log \left( \int_{b_{k-1}}^{b_k} N(x_i | \mu_i, \sigma_i^2) dx_i\right) \right).
	\end{gather}
	A model structure can be simply derived based on the above loss function (Figure \ref{fig:Figure3}A, red rectangle). However, the vanishing gradient problem was found when applying this model to estimate parameters in computer simulation, resulting in poor fitting for several variants. To deal with it, more detailed information should be taken into consideration to derive more accurate estimates. For this purpose, we focused on the overall fluorescence intensity distribution, which should be a mixture of Gaussians:
	
	\begin{gather}
	x \sim G(x|\bm{\pi},\bm{\mu},\bm{\sigma}^T\bm{\sigma}) = \sum_{i=1}^n \pi_i N(x|\mu_i,\sigma_i^2), \ i=1,2,...,n.
	\end{gather}
	
	Hence, the data generating process can be simulated using a reparameterized\cite{kingma2013auto} generator $G$ (Figure \ref{fig:Figure3}A, blue rectangle; Figure \ref{fig:Figure3}B), which specifically includes three steps: (1) sample an arbitaty variant $i$ from the categorical distribution $P(i|\bm{\pi})$ with probability $\pi_i$, which can be represented by $P(z=i|\bm{\pi}) = \pi_i$; (2) sample a random variable $u$ from a standard normal distribution, $u \sim N(u|0,1)$; (3) obtain the sample data of log-scaled fluorescence intensity by $\mu_i + \sigma_i u$. Moreover, to make the generator approximate to the true data distribution, a neural network-based discriminator $D$ was applied to determine whether the data is real or fake. In other words, a two-players game is played with value function $V(G,D)$:
	\begin{gather}
	\min \limits_G \max \limits_D V(G,D) = E_{x_{true} \sim p(x)}[\log (D(x_{true}))] + E_{x_{fake} \sim G(x|\bm{\pi},\bm{\mu},\bm{\sigma}^T \bm{\sigma})}[\log(1-D(x_{fake}))]
	\end{gather}
	
	Combine the above two parts togather, we can conclude the whole algorithm as below \ref{table:Table1}.
	
	\begin{table}[h]
		
		\label{table:Table1}
		
		\begin{tabular*}{16.5cm}{l}  
			
			\toprule
			\textbf{Algorithm 1} Parameter estimation model for FACS-seq. The default values are set as $t=8, \ m=640, \ \alpha_1=0.9$, \\ 
			$\alpha_2=0.1$. \\  
			\midrule   
			
			\quad \textbf{for} number of training iterations \textbf{do} \\  
			
			\quad \quad \textbf{for} $t$ steps \textbf{do}  \\
			
			\quad \vbox{\begin{itemize}
						\item Sample minibatch of $m$ samples $\{x_{fake}^{(1)}, x_{fake}^{(2)},..., x_{fake}^{(m)}\}$ from $G(x|\bm{\pi},\bm{\mu},\bm{\sigma}^T \bm{\sigma})$.
						\item Sample minibatch of $m$ samples $\{x_{true}^{(1)}, x_{true}^{(2)},..., x_{true}^{(m)}\}$ from real fluorescence intensity distribution.
						\item Update the discriminator by ascending its stochastic gradient: 
				  		\end{itemize}} \\
			
			\vbox{\begin{gather*} 
					\nabla_{\theta_d} \frac{1}{m} \sum_{j=1}^m [log (D(x_{true}^j)) + log(1-D(x_{fake}^j))]. 
				  \end{gather*}} \\
			
			\quad \quad \textbf{end for} \\
			
			\vbox {\begin{itemize}
					\item Sample minibatch of $m$ samples $\{x_{fake}^{(1)}, x_{fake}^{(2)},..., x_{fake}^{(m)}\}$ from $G(x|\bm{\pi},\bm{\mu},\bm{\sigma}^T \bm{\sigma})$.
					\item Update the generator by ascending its stochastic gradient:
				  \end{itemize}} \\
			
			\vbox{\begin{gather*} 
				\nabla_{\bm{\mu},\bm{\sigma}} \left[ \alpha_1 \frac{1}{n} \sum_{i=1}^n \sum_{k=1}^K \left( f_{ik} \log \left( \int_{b_{k-1}}^{b_k} N(x_i | \mu_i, \sigma_i^2) dx_i\right) \right) + \alpha_2 \frac{1}{m} \sum_{j=1}^m (log(D(x_{fake}^j))) \right]. 
				\end{gather*}} \\
			
			\quad \textbf{end for} \\
			
			\bottomrule  
			
		\end{tabular*}
		
	\end{table}
	
	\section{Results}
	\subsection{Simulation results}
	In order to evaluate the performance of the model, we first applied this model to estimate parameters in computer simulation. The data were generated via the following steps:
	\begin{itemize}
		\item Set $n=1,500, \ K=6$;
		\item $\mu_i \sim N(-0.08,0.4^2)$ and $-1.5<\mu_i<1, \ i=1,2,...n$; 
		\item $\sigma_i \sim N(-0.18,0.07^2)$ and $0.05<\sigma_i<0.4, \ i=1,2,...n$;
		\item $\log (\pi_i) \sim N(1,0.2^2)$, each $\pi_i$ was normalized by $1 / \sum_{i=1}^n \pi_i$;
		\item $x \sim \sum_{i=1}^n \pi_i N(\mu_i, \sigma_i^2)$;
		\item $\bm{b} = [-\infty, -0.8, -0.4, 0, 0.4, 0.8, \infty]$;
		\item $f_{i,k} = \int_{b_{k-1}}^{b_k} N(x_i|\mu_i, \sigma_i^2) dx_i, \ i=1,2,...n, \ k=1,2,...,K$.
	\end{itemize}
	
\begin{figure}[h]
		\centering
		\includegraphics[scale=1.3]{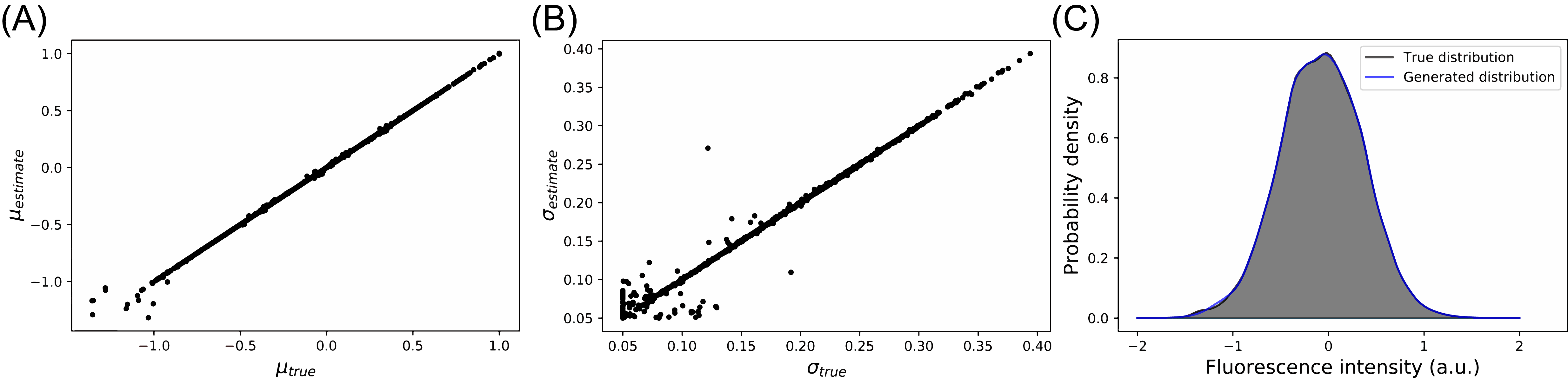}
		\caption{Simulation results suggest the model is reliable to estimate parameters for gene expression distribution. (A, B) The agreement between estimates and real parameters is very good for both (A) mean ($r^2=0.9986, \ n=1,500$) and (B) standard deviation ($r^2=0.9862, \ n=1,500$). (C) The generated distribution is approx to the real data distribution.}
		\label{fig:Figure4}
	\end{figure}

	We then fed the above information exclude $\bm{\mu}$ and $\bm{\sigma}$ into our model. As a result, the agreement between the estimates and real parameters is very good for both mean (Figure Figure \ref{fig:Figure4}A) and standard deviation (Figure Figure \ref{fig:Figure4}B).
	
	\subsection{Case study: FACS-seq-based profiling of \emph{tnaC}-mediated tryptophan-dependent gene expression systems}
	As a molecular sensor, \emph{tnaC} encodes a 24-residue leader peptide that responds to intracellular tryptophan as well as regulates the biosynthesis of indole. Wang et al.\cite{wang2020dynamics} constructed a comprehensive codon-level mutagenesis library of \emph{tnaC} based on a two-color reporter system in \emph{E. coli}, where the \emph{egfp} is under the control of \emph{tnaC} as the sensor response reporter, and \emph{mcherry} is constitutively expressed to normalize cell-to-cell variation. Cells treated with a particular concentration of ligand were sorted into six bins according to their responses, the read number of each bin was quantified through NGS. As a proof of the concept, we tested our model on their dataset.
	
	NGS clean data were obtained from Bioproject \href{https://www.ncbi.nlm.nih.gov/sra/?term=PRJNA503322}{PRJNA503322} of NCBI Sequence Read Archive. Pairs of paired-end data were merged by FLASH script and those reads without detected pairs were removed. The regular expression pattern of 'TAAGTGCATT[ATCG]\{70,80\}TTTGCCCTTC' in the sequencing reads (and the reverse complementary sequence) was searched, and those carrying mutations within the upstream (TAAGTGCATT) or downstream (TTTGCCCTTC) flanking regions (10 bp each) were removed, the read number of each distinct sequence was counted. 
	
	For the two-color report system, cells were sorted based on the ratio between the log-scaled fluorescence intensity of \emph{eGFP} and \emph{mCherry}, Suppose a set of boundaries $\{eGFP=a\cdot mCherry+b_j | j=1,...,K-1\}$ were used to sort the cells into $K$ bins, we can define $x=eGFP/mCherry^a$ to represent the expression of each cell, thus the boundaries corresponding to $x$ should be $\bm{b}=\{b_0=-\infty,b_1,...,b_{K-1},b_K=\infty\}$. As for the NGS data of each bin, the read counts were first normalized by a factor $C_k/R_k$, where $C_k$ is the total number of cells sorted in $bin_k$ and $R_k$ is the total read counts in $bin_k$, these values were then normalized across all bins to derive the frequency distribution of each variant.
	
\begin{figure}[h]
		\centering
		\includegraphics[scale=1.3]{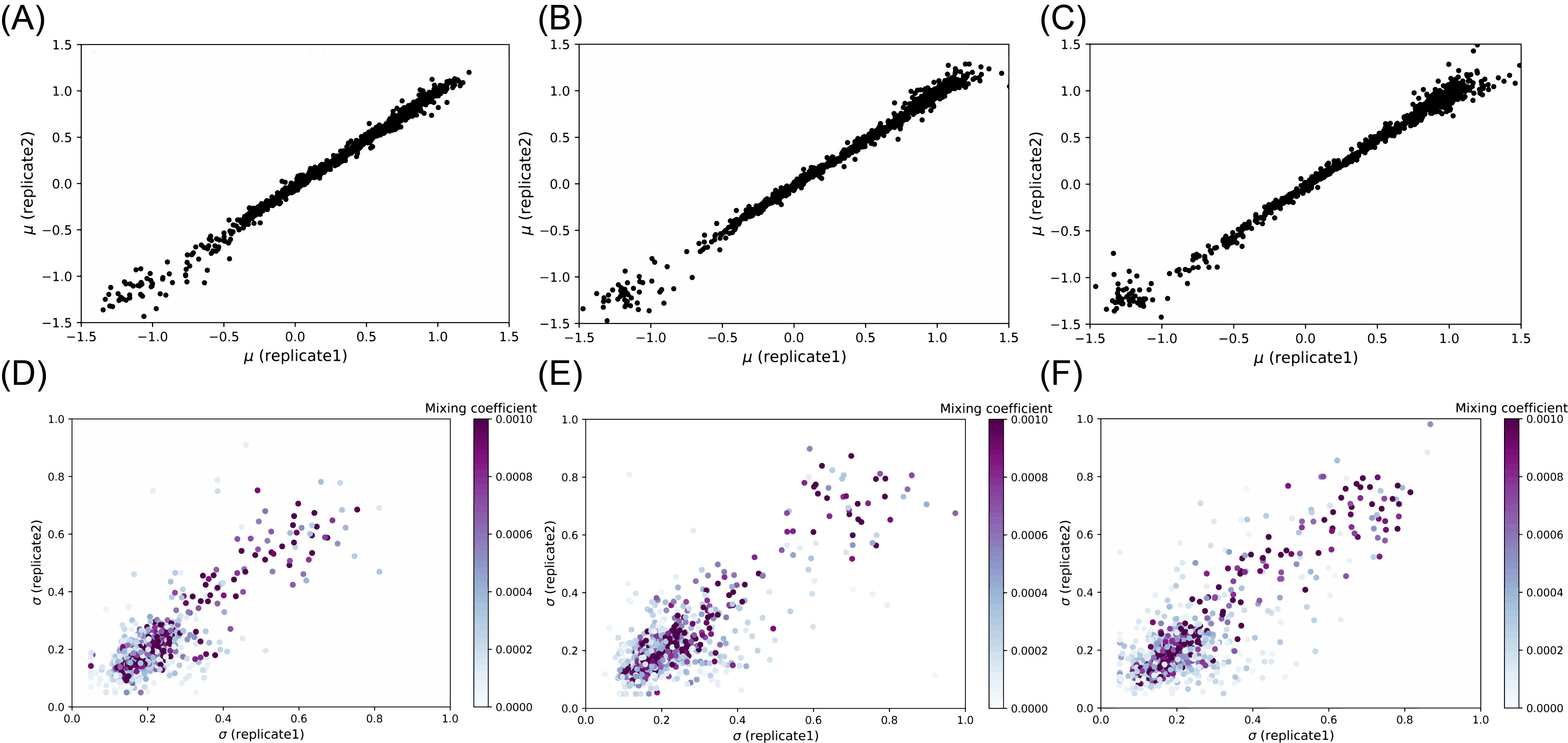}
		\caption{Model performances in experimental dataset demonstrated the power of the model to capture gene expression levels as well as variations. (A, B, C) Expression levels revealed strong consistency between biological replicates for all three experimental conditions (A: 0 $\mu M$, Pearson corrlation coefficient (PCC) = 0.9954, n=1,367; B: 100 $\mu M$, PCC = 0.9960, n=1,367; C: 500 $\mu M$, PCC = 0.9938, n=1,367.). (D, E, F) Expression variations showed good correlations between replicates (D: 0 $\mu M$, PCC = 0.8299, n=1,367; E: 100 $\mu M$, PCC = 0.8160, n=1,367; F: 500 $\mu M$, PCC = 0.8394, n=1,367.).}
		\label{fig:Figure5}
	\end{figure}

	After the above pretreatment, we got three datasets corresponding to three different concentrations of ligand supplementation (0, 100, 500 $\mu M$). We then performed our method on these datasets, as a result, the correlations between replicates proved the ability of our model that can capture the properties of gene expression of each individual in a pooled library (Figure \ref{fig:Figure5} A-F). However, the model performances in predicting the standard deviation were not so good compared to simulation results, which we thought were due to the observational errors during experiments. These errors may include polymerase preference, the random fluctuation in testing fluorescence intensity, the misclassification when sorting cells, etc. We further discovered that those errors are more likely to influence the variants with small mixing coeffecient (\ref{fig:Figure5} D-F), this phenomenon is obvious since few measurements of them would exacerbate the statistical uncertainty.
	
	\section{Conclusion}
	In this paper, we proposed a novel method that can precisely estimate the parameters of gene expression distribution based on the data derived from FACS-seq experiments. This model makes full use of all observation data as well as the assumption that gene expression follows a log-normal distribution, which to the author's best knowledge, is the first effective model that enables high-throughput quantification of phenotypic heterogeneity. We believe this model can help to facilitate the research progress of corresponding fields.
	
	In this model, we only focus on the ideal situation because we don't have any prior knowledge of these observational errors. A more robust model can be constructed if we can measure the distribution of these errors, or we can reduce them in our experiment to improve data quality to achieve better performances. To derive more reliable parameter estimates, we recommend ensuring the uniformity of the library as those results associated with small proportions are more likely to be affected by the experimental noise. Besides, more bins are recommended when sorting cells as they can produce more information.
	
	\section{Data and code availability}
	The code and data related to this work can be accessed via Github (\href{https://github.com/fenghuibao/Deep_FS.git}{https://github.com/fenghuibao/Deep\_FS.git}).
	
	
		
		
	\pagebreak
	\medskip
	
	\small
	\bibliographystyle{unsrt}
	\bibliography{Reference.bib}
	
\end{document}